\documentstyle[preprint,eqsecnum,aps]{revtex}

\begin{document}
\input epsf

\count255=\time\divide\count255 by 60 \xdef\hourmin{\number\count255}
  \multiply\count255 by-60\advance\count255 by\time
  \xdef\hourmin{\hourmin:\ifnum\count255<10 0\fi\the\count255}

\draft
\preprint{WM-00-109}

\title{Bounds on Bosonic Topcolor}

\author{Alfredo Aranda\footnote{aranda@physics.wm.edu}
 and  Christopher D. Carone\footnote{carone@physics.wm.edu}}

\vskip 0.1in

\address{Nuclear and Particle Theory Group, Department of
Physics, College of William and Mary, Williamsburg, VA 23187-8795}

\vskip .1in
%\date{\hourmin \ \today}
\date{July, 2000}
\vskip .1in

\maketitle
%\tightenlines

\begin{abstract}
We consider the phenomenology of models in which electroweak symmetry 
breaking is triggered by new strong dynamics affecting the third generation 
and is transmitted to the light fermions via a fundamental Higgs doublet.  
While similar in spirit to the old bosonic technicolor idea, such `bosonic
topcolor' models are allowed by current phenomenological constraints,
and may arise naturally in models with large extra dimensions.  We study 
the parameter space of a minimal low-energy theory, including bounds 
from Higgs boson searches, precision electroweak parameters, and flavor 
changing neutral current processes.  We show that the model can provide 
a contribution to $D^0$-$\overline{D^0}$ mixing as large as the current 
experimental bound.
\end{abstract}

\thispagestyle{empty}

\newpage
\setcounter{page}{1}

\section{Introduction} \label{sec:intro}
In spite of the quantitative success of the standard model, the 
mechanism of electroweak symmetry breaking remains unclear.  Only a few 
years ago, bosonic technicolor models provided a relatively unconventional 
approach to solving this problem~\cite{c1,c2,c3}: electroweak symmetry 
was broken dynamically by a fermion condensate triggered by new strong 
forces, while a fundamental scalar field was responsible for transmitting 
these effects to the fermions through ordinary Yukawa couplings.  These 
models did not require a conventional extended technicolor sector, and 
hence were freed from the associated flavor changing neutral current (FCNC)
problems.  Unfortunately, precision electroweak constraints rule out 
bosonic technicolor models at least in models where the strong dynamics 
is QCD-like and the S-parameter can be reliably estimated~\cite{c4}.

In this letter, we point out that a very similar scenario, bosonic topcolor,
also provides a very simple low-energy effective theory, but one that is 
not in conflict with electroweak constraints.  In this scenario, 
electroweak symmetry is partly broken by new strong dynamics that affects 
fields of the third generation, as in conventional topcolor 
scenarios~\cite{c5,c6}, while a weakly-coupled scalar doublet transmits 
the symmetry breaking to the fermions via Yukawa couplings.  Since this 
scenario involves both a fundamental ($H$) and a composite ($\Sigma$) Higgs 
field that both contribute to electroweak symmetry breaking, the usual 
problematic relation~\cite{c5} between the dynamical top quark mass and 
the electroweak symmetry breaking scale is not obtained.  The result is a 
viable two Higgs doublet model of type III, which we will show survives 
the bounds from flavor changing neutral current processes and may provide 
interesting flavor-changing signals as well.

The possibility of topcolor models involving fundamental scalars has been 
considered in Refs.~\cite{c7,c8}.  In these papers, however, the 
fundamental Higgs field was strongly coupled, and the authors considered 
whether the fundamental field itself could trigger the formation of 
a $t\overline{t}$ condensate.  Here we introduce $H$ as a weakly-coupled 
field and investigate the phenomenological consequences.

It is worth pointing out that a philosophical objection to the original 
bosonic technicolor scenarios, and the bosonic topcolor models described 
here, is that strong dynamics was originally intended to eliminate the need 
for a fundamental Higgs field altogether, as well as the associated problem 
with quadratic divergences.  Recent theoretical developments relating to the 
possibility of low-scale quantum gravity~\cite{c9} renders these 
objections hollow:  The presence of a low string scale eliminates the 
conventional desert so that nonsupersymmetric low-energy theories with 
fundamental scalars are not unnatural.  Moreover, in this setting there 
are new origins for the strong dynamics, namely the exchange of a 
nonperturbatively large number of gluon Kaluza-Klein excitations~\cite{c10}.  
While we will not consider an explicit extra-dimensional embedding 
of the scenario described here, it seems that these considerations make the 
investigation of models with dynamical electroweak symmetry breaking and 
fundamental scalar fields well motivated.

In the next section we will present a simple realization of the bosonic 
topcolor idea following the Nambu-Jona-Lasinio approach~\cite{c5}.  Our first
model is non-generic in the sense that we do not specify the most general 
set of higher-dimension operators that could appear in an arbitrary 
high-energy theory.  However, it does provide a very convenient framework 
for parameterizing and exploring the basic phenomenological features of 
the scenario.  After considering the phenomenological bounds, we will 
describe how to study the same type of scenario in a more general effective 
field theory approach.  While we will not consider every phenomenological 
detail in this letter, we hope to obtain an accurate overall picture of 
the allowed parameter space.  Finally, we will discuss flavor changing 
signals for the model, notably a potential contribution to 
$D^0$-$\overline{D^0}$ mixing that can be as large as the current 
experimental bound.  We then summarize our conclusions.

\section{Minimal Bosonic Topcolor} \label{sec:mbtc}

Our high-energy theory is defined by
\begin{equation}
{\cal L} = {\cal L}_{\rm H} + {\cal L}_{\rm NJL} \,\,\, ,
\end{equation}
where
\begin{equation}
{\cal L}_{\rm H} = D_\mu H^\dagger D^\mu H - m^2_H H^\dagger H
- \lambda (H^\dagger H)^2 - h_t (\overline{\psi}_L H t_R + h.c.)  \,\,\, ,
\label{eq:h}
\end{equation}
and
\begin{equation}
{\cal L}_{\rm NJL} 
= \frac{\kappa}{\Lambda^2} \overline{\psi}_L t_R \overline{t}_R \psi_L \,\,\, .
\label{eq:njl}
\end{equation}
The field $H$ is a fundamental scalar doublet, and $\Lambda$ characterizes the 
scale at which new physics is present that generates the nonrenormalizable
interaction in Eq.~(\ref{eq:njl}).  In light of our introductory remarks,
we will assume henceforth that $\Lambda \mbox{\raisebox{-1.0ex} 
{$\stackrel{\textstyle ~<~}{\textstyle \sim}$}} 100$~TeV.  In this minimal 
scenario we assume that the right-handed top, and left-handed top-bottom 
doublet $\psi_L$ experience the new strong interactions.  Immediately 
beneath the scale $\Lambda$ we may rewrite Eqs.~(\ref{eq:h}) and 
(\ref{eq:njl}) as \begin{eqnarray} \label{lagrangian}
\nonumber
{\cal{L}} & = & D_{\mu} H^{\dagger} D^{\mu} H 
- m^{2} H^{\dagger} H  - \lambda \left( 
H^{\dagger} H \right)^{2} - c \Lambda^{2} \Sigma^{\dagger} \Sigma
 \\
          &   & - h_{t} ( \bar{\Psi}_{L} t_{R} H + h.c.) -
g_t (\bar{\Psi}_{L} t_{R} \Sigma + h.c.)
\end{eqnarray}
where $\Sigma$ is a non-propagating auxiliary field.  Using the equations of
motion, $\Sigma =-g_t (\overline{t}_R \psi_L) / (c \Lambda^2)$ and one recovers
Eqs.~(\ref{eq:h}) and (\ref{eq:njl}) with the identification 
$\kappa=g_t^2/c$.

At energies $\mu \ll \Lambda$, quantum corrections induce a kinetic term for 
$\Sigma$, so that it becomes a dynamical field, a composite Higgs doublet 
in the low-energy theory.  In order to study the quantum corrections to 
Eq.~(\ref{lagrangian}) it is convenient for us to define the column vector
\begin{eqnarray} \label{phi}
\Phi = \left( \begin{array}{c} \Sigma \\ H \end{array} \right) \, .
\end{eqnarray} 
Then the kinetic term at the scale $\mu$ may be written  $ \partial_{\mu}
\Phi^{\dagger} Z \partial^{\mu} \Phi$, with
\begin{eqnarray} \label{Z}
Z = \left( \begin{array}{cc}
\frac{g_t^2 N_{C} \ln(\Lambda/\mu)}{8 \pi^{2}} & 
\frac{g_t h_{t}N_{C}\ln(\Lambda/\mu)}{8 \pi^{2}} \\
\frac{g_t h_{t}N_{C}\ln(\Lambda/\mu)}{8 \pi^{2}} &
1 + \frac{h_{t}^{2} N_{C} \ln(\Lambda/\mu)}{8 \pi^{2}}
\end{array} \right) 
\approx
\left( \begin{array}{cc}
\frac{1}{r^{2}} & 
0 \\ 0 & 1 \end{array} \right) \, .
\end{eqnarray}
In general, $Z$ must be diagonalized and rescaled so that the kinetic terms 
assume the canonical form.  However, in most of the parameter space that we
consider later in this paper $h_t$ is small enough that the off-diagonal 
elements of $Z$ are numerically irrelevant;  thus we use the simpler 
approximate form parameterized by $r$ in Eq.~(\ref{Z}).  Properly normalized 
kinetic terms are then obtained via the substitution 
$\Sigma \rightarrow r\Sigma$.  Quantum corrections also induce quartic
self interactions, and mixing in the $\Phi$ mass matrix.  We retain the
largest self-coupling, $(\Sigma^\dagger \Sigma)^2$ with coefficient 
$\lambda_{\Sigma} = g_t^4 N_{C} \ln(\Lambda/\mu)/(4 \pi^2)$; the $\Phi$
mass matrix is given by
\begin{eqnarray} \label{lmass}
{\cal L}_{mass} = - \Phi^{\dagger} {\cal M}^2 \Phi \, ,
\end{eqnarray}
with
\begin{eqnarray} \label{M}
{\cal M}^{2} = \left( \begin{array}{cc}
r^2 m_\Sigma^2 & r \delta m^2 \\ r \delta m^2 & m_H^2 \end{array} \right) \, ,
\end{eqnarray}
where $\delta m^2 = -\frac{N_c}{8 \pi^2}g_t h_t \Lambda^2$.  Eq.~(\ref{M})
reflects the fact that both the diagonal and off-diagonal entries receive
quadratically divergent radiative corrections.  For the diagonal elements,
the tree-level mass terms present in Eq.~(\ref{lagrangian}) can be fine-tuned
against these radiative corrections (as in the standard model) so that 
$m_\Sigma$ and $m_H$ are well beneath the cutoff scale $\Lambda$.  On the 
other hand, there is no tree-level $H\Sigma$ mass mixing term given the 
way we defined our high-energy theory in Eqs.~(\ref{eq:h})-(\ref{eq:njl}).  
However, since we are considering the situation where the scale $\Lambda$ is 
relatively low ($<100$~TeV) and where the coupling $h_t$ is small 
(see Figs.~1 and 2), the off-diagonal elements will also be much smaller 
than the cut off.  For the case where such tree-level mass mixing is 
present at the scale $\Lambda$, the reader should refer to Section~3.

Electroweak symmetry will be broken in this model if $\Sigma$ and/or $H$ 
acquire vacuum expectation values (vevs).  There are several ways this can 
happen depending on the values of the different parameters in the model.  
We are principally interested in the case where $m_H^2 > 0$, so that 
electroweak symmetry breaking is triggered by the strong dynamics and 
the vev of $H$ can be interpreted as a subsidiary effect.  Thus,
it is necessary to study the scalar potential,
\begin{eqnarray} \label{potential}
\nonumber
V(\Sigma,H) & = & r^2 m_\Sigma^2 \Sigma^{\dagger} \Sigma + m_H^2 H^{\dagger}
H + r \delta m^2 \left( \Sigma^{\dagger} H + h.c. \right) \\
            &   & + \lambda \left( H^{\dagger} H\right)^2 +
\lambda_{\Sigma} r^4 \left( \Sigma^{\dagger} \Sigma \right)^{2} \, .
\end{eqnarray}
Rather than search directly for minima in a five-dimensional parameter space 
($m^2_\Sigma$, $m^2_H$, $\delta m^2$, $\lambda$, $\lambda_\Sigma$) we 
extremize the potential and solve for $m_\Sigma^2$ and $m_H^2$ in terms of 
the $\Sigma$ and $H$ vevs.  It is much more manageable to study the 
remaining constrained three-dimensional parameter space and determine 
which points correspond to stable local minima.  If we denote the vevs 
of $\Sigma$ and $H$ by $v_1/\sqrt{2}$ and
$v_2/\sqrt{2}$, one finds
\begin{eqnarray}\label{m1m2}
m_\Sigma^2 & = & - \frac{1}{ r v_{1}}\left( \delta m^2 v_2 + 
\lambda_{\Sigma} r^3 v_{1}^3 \right)\, ,\\
m_H^2 & = & - \frac{1}{ v_2}\left(\delta m^2 r v_{1} +  \lambda v_2^3  \right)\, .
\end{eqnarray}
From Eq.~(\ref{potential}), one may obtain the mass matrices for the scalars,
pseudoscalars, and charged scalars:
\begin{eqnarray} \label{scalar}
M_S = \frac{1}{2} \left( \begin{array}{cc}
m_\Sigma^2 r^2 + 3 \lambda_{\Sigma} r^4 v_1^2 & \delta m^2 r \\
\delta m^2 r & m_H^2 + 3 \lambda v_2^2
\end{array} \right) \, ,
\end{eqnarray}
\begin{eqnarray} \label{pseudo}
M_P = \frac{1}{2} \left( \begin{array}{cc}
m_\Sigma^2 r^2 +  \lambda_{\Sigma} r^4 v_{1}^2 & \delta m^2 r \\
\delta m^2 r &m_H^2 +  \lambda v_2^2
\end{array} \right) \, ,
\end{eqnarray}
\begin{eqnarray} \label{charged}
M_{+} =  \left( \begin{array}{cc}
 m_\Sigma^2 r^2 + \lambda_{\Sigma} r^4 v_{1}^2 &  \delta m^2 r \\
 \delta m^2 r &  m_H^2 + \lambda  v_2^2
\end{array} \right) \, .
\label{eq:mplus}
\end{eqnarray}
The Higgs field vevs are responsible for producing the proper gauge
boson masses, {\em i.e.}
\begin{eqnarray} \label{ew-relation}
v_1^2 + v_{2}^2        & = & (246 \,\,\rm{GeV})^2 \, ,
\end{eqnarray}
as well as the mass of the top quark
\begin{equation}
m_t = (r g_t v_1 + h_t v_2)/\sqrt{2}  \,\,\, .
\label{eq:top}
\end{equation}
This expression shows that the top quark receives both an ordinary
and a dynamical contribution.  Since we focus on small values of $h_t$
in this letter, the top quark mass is mostly dynamical, originating
from the first term in Eq.~(\ref{eq:top}).  In this limit, the vevs
$v_1$ and $v_2$ are determined by the choice of scales $\Lambda$ and
$\mu$, since the quantity $rg_t$ is independent of $g_t$.

\section{Phenomenology}

Notice that all the freedom in Eqs.~(\ref{m1m2})-(\ref{eq:mplus}) is 
fixed by specifying $\Lambda$, $\mu$, $h_t$ and $\lambda$.  Thus, for 
a fixed choice of $\Lambda < 100$~TeV and $\mu$ of order the weak scale, 
we may map our results onto the $\lambda$-$h_t$ plane.  Fig.~1 displays 
results for $\Lambda=10$~TeV and Fig.~2 for $\Lambda=100$~TeV, with 
$g_t=1$.  In each case, the intersecting solid lines indicate 
where $m_\Sigma^2$ or $m_H^2$ change sign, with positive values lying 
above the corresponding line. Figs.~1a and 2a provide mass contours for 
the lightest neutral scalar and charged scalar states; Figs. 1b and 2b 
display constant contours for the electroweak parameters $S$ and $T$.  
These were computed using formulae available in the literature for general 
two Higgs doublet models~\cite{c11},
\begin{eqnarray} \label{S}
\nonumber
S  & = & \frac{1}{12 \pi} \left( s_{\alpha-\beta}^2
\left[ \ln\frac{M_2^2}{M_H^2} + g(M_1^2,M_3^2)-\frac{1}{2} 
\ln\frac{M_+^2}{M_1^2}-\frac{1}{2} \ln\frac{M_+^2}{M_3^2}\right] \right. \\
  &   &  + \left. c_{\alpha-\beta}^2 \left[\ln\frac{M_1^2}{M_H^2} + 
g(M_2^2,M_3^2)-\frac{1}{2} \ln\frac{M_+^2}{M_2^2}-
\frac{1}{2} \ln\frac{M_+^2}{M_3^2} \right] \right) \, ,
\end{eqnarray}
and
\begin{eqnarray} \label{T}
\nonumber
T  & = & \frac{3}{48 \pi s^2 m_W^2} \left( s_{\alpha-\beta}^2
\left[ f(M_1^2,M_+^2)+f(M_3^2,M_+^2)-f(M_1^2,M_3^2) \right] \right. \\
  &   &  + \left. c_{\alpha-\beta}^2 \left[f(M_2^2,M_+^2)+
f(M_3^2,M_+^2)-f(M_2^2,M_3^2) \right] \right) \, ,
\end{eqnarray}
where $M_1$, $M_2$, $M_3$, and $M_+$ are the light scalar, heavy
scalar, pseudoscalar, and charged scalar masses respectively,
and $\beta=\tan^{-1}(v_2/v_1)$. The scalar mixing angle $\alpha$ and 
the functions $f$ and $g$ are defined in Ref.~\cite{c11}.  Figs.~1c 
and 2c show regions excluded by (i) the current LEP bound on neutral
Higgs production, (ii) bounds on the $S$ and $T$ parameters,
(iii) bounds on the charged scalar mass from $b\rightarrow s \gamma$.
In the first case, we compute the production cross section for the
lightest scalar state $\phi_s$,
\begin{equation}
\sigma(e^+ e^- \rightarrow Z \phi_s) = s^2_{\alpha-\beta}
\sigma_{SM}(e^+ e^- \rightarrow Z H^0)  
\end{equation} 
and compare to the corresponding standard model cross section for
a Higgs boson with mass equal to the current LEP bound, $m_H < 107.9$~GeV,
95\% CL~\cite{c12}.  In the case of the $S$ and $T$ bounds, we consider 
the results of global electroweak fits quoted in the Review of 
Particle Properties~\cite{c13}, $S=-0.26 \pm 0.14$ and 
$T=-0.11 \pm 0.16$~\cite{c4}, which assume a reference Higgs mass of 300~GeV.  
We show the two standard deviation limit contours for $S$
and $T$ separately wherever an upper or lower limit is exceeded. (Note that
we don't take into account correlations between $S$ and $T$ in determining
this exclusion region.)  Finally, we plot the charged Higgs mass limit
$m_{H^+}>[244+63/(\tan\beta)^{1.3}]$~GeV from 
$b\rightarrow s \gamma$~\cite{c14}.  This is the strongest, albeit
indirect, charged Higgs mass limit listed in Ref.~\cite{c13}. Although, 
strictly speaking, this bound applies to a type II two-Higgs doublet 
model, the leading top quark loop contribution is the same in
our model; the top quark-charged scalar coupling is given by
\begin{eqnarray}
\Phi^+ \frac{g}{2\sqrt{2}M_W}& & \left[ \overline{t} 
( m_t^H \cot\beta - m_t^\Sigma \tan\beta) V_{tq} (1-\gamma^5) q \right. 
\nonumber \\  
&-& \left. \overline{t} \cot\beta V_{tq} m^H_q (1+\gamma^5) 
q\right]
\label{eq:chint}
\end{eqnarray}
in the case where the Cabibbo-Kobayashi-Maskawa (CKM) matrix $V$ originates 
from diagonalization of the down quark Yukawa matrix alone (the reason for 
this assumption is given in the following section).  Here $m^H_t$ and 
$m^\Sigma_t$ refer to contributions to the top mass from the $H$ and 
$\Sigma$ vevs, respectively. For most of the parameter range of
interest to us, $m_t^H \ll m_t^\Sigma$ and the interaction in 
Eq.~(\ref{eq:chint}) reduces to that of a type II model, and the 
$b \rightarrow s \gamma$ bound is approximately valid.  In both Figs.~1 and 2
a  rectangular region is shown in which the charged scalars are
heavy enough to weaken the flavor changing neutral current bounds, 
without exceeding that of the $T$ parameter.

\section{Flavor Changing Signals} \label{sec:fcs}

The fact that one of our two Higgs doublets ($\Sigma$) couples 
preferentially to the top quark leads to a potentially interesting source 
of flavor violation in the model.  While the charge $-1/3$ quark masses 
and neutral scalar interactions both originate via couplings to $H$ (and 
hence are simultaneously diagonalizable), the same is not true in the 
charge $2/3$ sector, where the mass matrix depends on both the $H$ and 
$\Sigma$ vevs,
\begin{equation}
M^U = Y^U_\Sigma \frac{v_1}{\sqrt{2}} +  Y^U_H \frac{v_2}{\sqrt{2}} \,\,\, .
\end{equation}
For concreteness, let us consider a definite Yukawa texture:
\begin{equation}
M^U = \left(\begin{array}{ccc}0 &0&0\\ 0 & 0 & 0 \\ 0 & 0 & r g_t \end{array}
\right)\frac{v_1}{\sqrt{2}} + \left(\begin{array}{ccc} \lambda^8 & \lambda^5 &
\lambda^3 \\ \lambda^5 & \lambda^4 & \lambda^2 \\ \lambda^3 & \lambda^2 & h_t
\end{array}\right)\frac{v_2}{\sqrt{2}} \,\,\, .
\end{equation}
Here $\lambda=0.22$ is the Cabibbo angle, and we have picked a symmetric 
texture for the fundamental Higgs Yukawa couplings that approximately 
reproduces the correct CKM angles.  Dropping the 
factors of $v_i/\sqrt{2}$, the matrices shown give the neutral scalar
couplings in our original field basis.  In the quark (and Higgs) mass 
eigenstate basis, there will be flavor-changing top and charm quark 
interactions.  Here, we focus only on the latter.  CKM-like rotations 
that diagonalize the mass matrices yield $1$-$2$ neutral scalar couplings 
of order $\lambda^5$.  It is straightforward to estimate the contribution 
to $D^0$-$\overline{D^0}$ mixing,
\begin{equation}
|\frac{\Delta m_D}{m_D}|_{{\rm new}} 
\approx \lambda^{10} \frac{f_D^2}{12 M_\phi^2}\left[
-1+11\frac{m_D^2}{(m_c+m_u)^2}\right]  \,\,\, .
\label{eq:ddbar}
\end{equation}
For $f_D\approx 200$~MeV, this contribution saturates the current experimental
bound, $\Delta m_D < 1.58\times 10^{-10}$~MeV~\cite{c13}, for 
$M_\phi \mbox{\raisebox{-1.0ex} {$\stackrel{\textstyle ~<~}{\textstyle \sim}$}}
495$~GeV. The reason that we do not include this as a bound is that the 
$1$-$2$ neutral scalar couplings need not be $O(\lambda^5)$; they could in 
principle be zero, if the CKM matrix results from the diagonalization of the 
down quark Yukawa matrix alone.  Since this is the least constrained
possibility, we adopted this assumption in Eq.~(\ref{eq:chint}) for
computing the $b\rightarrow s \gamma$ exclusion region.  Generically, 
however, we see that bosonic topcolor models predict significant 
contributions to $D^0$-$\overline{D^0}$ mixing, potentially as large 
as the current experimental bound.

\section{Generalizations} \label{sec:gen}

The scenario described in the previous section is particularly convenient 
in that the basic phenomenology can be described in a two-dimensional 
parameter space, for fixed $\Lambda$ and $\mu$.  However, a realistic 
high-energy theory is likely to provide more than the single higher-dimension 
operator in Eq.~(\ref{eq:njl}).  In this section we briefly describe the 
effective field theory approach for constructing the most general 
low-energy effective bosonic topcolor theory. Given our assumption
that $\psi_L$ and $t_R$ experience the new strong dynamics, the 
strongly-coupled sector of the theory possesses a global symmetry 
$G=$SU(2)$_L\times$U(1)$_R$, that is spontaneously broken by the 
$t\overline{t}$ condensate to the U(1) that counts top quark number.  If we 
denote the elements of this SU(2)$\times$U(1) by $U$ and $V$, respectively, 
then the transformation properties of the fields are given by 
\begin{equation}
\psi_L \rightarrow U \psi_L ,  \,\,\,\,\,\,\,\,\,\,
t_R \rightarrow V t_R , \,\,\,\,\,\,\,\,\,\, \mbox{ and } 
\,\,\,\,\,\,\,\,\,\, \Sigma \rightarrow U \Sigma V^\dagger \,\,\, ,
\end{equation}
where $V$ is a phase.  The Yukawa couplings of the fundamental Higgs field 
explicitly break $G$, so we may treat $h_t H$ as a `spurion' transforming
as
\begin{equation}
h_t H \rightarrow U (h_t H) V^\dagger  \,\,\, .
\end{equation}
We may now include $h_t H$ systematically in an effective Lagrangian 
by forming all possible $G$-invariant terms.  At the renormalizable
level,
\begin{eqnarray}
{\cal L}_{eff}&=& D^\mu H^\dagger D_\mu H + D^\mu \Sigma^\dagger 
D_\mu \Sigma \nonumber\\
&-& m_H^2 H^\dagger H - m_\Sigma^2 \Sigma^\dagger\Sigma 
+ m^2_{H\Sigma}h_t (H^\dagger \Sigma +h.c.) \nonumber \\
&-&\lambda (H^\dagger H)^2 - \lambda_0 (\Sigma^\dagger \Sigma)^2
+h_t (h^\dagger \Sigma)\Sigma^\dagger \Sigma +\cdots \nonumber \\
&-&h_t \overline{\psi}_L H t_R - g_\Sigma \overline{\psi}_L \Sigma t_R 
+h.c. \,\,\, .
\label{eq:new}
\end{eqnarray}
Note that we have eliminated a possible kinetic mixing term by field 
redefinitions, which do not affect the form of the other terms.  
The $\cdots$ represent all the other possible quartic terms which are 
higher order in $h_t$.  Unlike the model described in the previous section, 
we no longer have a boundary condition at the scale $\Lambda$ that 
sets $\lambda_0(\Lambda)=0$ and $m^2_{H\Sigma}(\Lambda)=0$, thus
introducing two additional degrees of freedom into the scalar potential.  Since
we are now working directly with the low-energy theory, the scale $\Lambda$ is
not input directly, but rather can be computed by determining the scale at 
which the wavefunction renormalization of the $\Sigma$ field vanishes.  
At this scale, $\Sigma$ again becomes an auxiliary field, and may be 
eliminated using the equations of motion, leaving a more general set of 
higher-dimension operators than we had assumed originally 
in Eq.~(\ref{eq:njl}).

A complete investigation of the parameter space of this generalized model is
beyond the scope of this letter.  Before closing, we point out that there are
reasonable parameter choices in Eq.~(\ref{eq:new}) where the resulting
phenomenology is similar to the minimal model considered in Section~2.  
In Fig.~3 we provide the same information given in Figs.~1 and 2 for the 
general bosonic topcolor model, with $m^2_{H\Sigma}=(400 \mbox{ GeV})^2$ 
and $\lambda_0=1$.  It is interesting that in this case the allowed band 
delimited by the FCNC and $T$ parameter lines lies mostly in the region 
where both $m^2_\Sigma$ and $m^2_H$ are positive; in this region the mixing 
term in Eq.~(\ref{M}) drives one of the scalar mass squared eigenvalues 
negative so that electroweak symmetry is broken.  A full exploration of 
this parameter space will be provided elsewhere~\cite{aranda}.

\section{Conclusions} \label{sec:concl}

In this letter, we have described models in which electroweak symmetry breaking
is triggered by strong dynamics affecting the third generation but transmitted
to the fermions by a weakly-coupled, fundamental Higgs doublet. We have 
argued in the Introduction that such models are not unnatural given recent 
developments in low-scale quantum gravity.  Our minimal scenario, while 
probably not representing the ultimate high-energy theory, has the virtue of 
allowing a simple parameterization of the basic phenomenology of the model.  
It is our hope that others will adopt it as the basis for further 
phenomenological study.  Issues that one could address include relaxation
of our small $h_t$ approximation, flavor-changing top quark processes, and
the collider physics of the model. We also described how the scenario may be
generalized using effective field theory techniques.  Unlike bosonic 
technicolor models, bosonic topcolor is not excluded by current 
phenomenological bounds.  Moreover, the model has interesting flavor-changing 
signals such as a contribution to $D^0$-$\overline{D^0}$ mixing that could 
be as large as current experimental bounds.

\begin{center}
{\bf Acknowledgments}
\end{center}
We thank the National Science Foundation for support under Grant Nos.\ 
PHY-9800741 and PHY-9900657, and the Jeffress Memorial Trust for support under 
Grant No. J-532.

%\appendix

%%%%%%%%%%%%%%%%%%%%%%%%%%%%%%%%%%%%%%%%%%%%%%%%%%%%%
%   FIGURE
%%%%%%%%%%%%%%%%%%%%%%%%%%%%%%%%%%%%%%%%%%%%%%%%%%%%%
\begin{figure}  
\centerline{\epsfysize=2.75 in \epsfbox{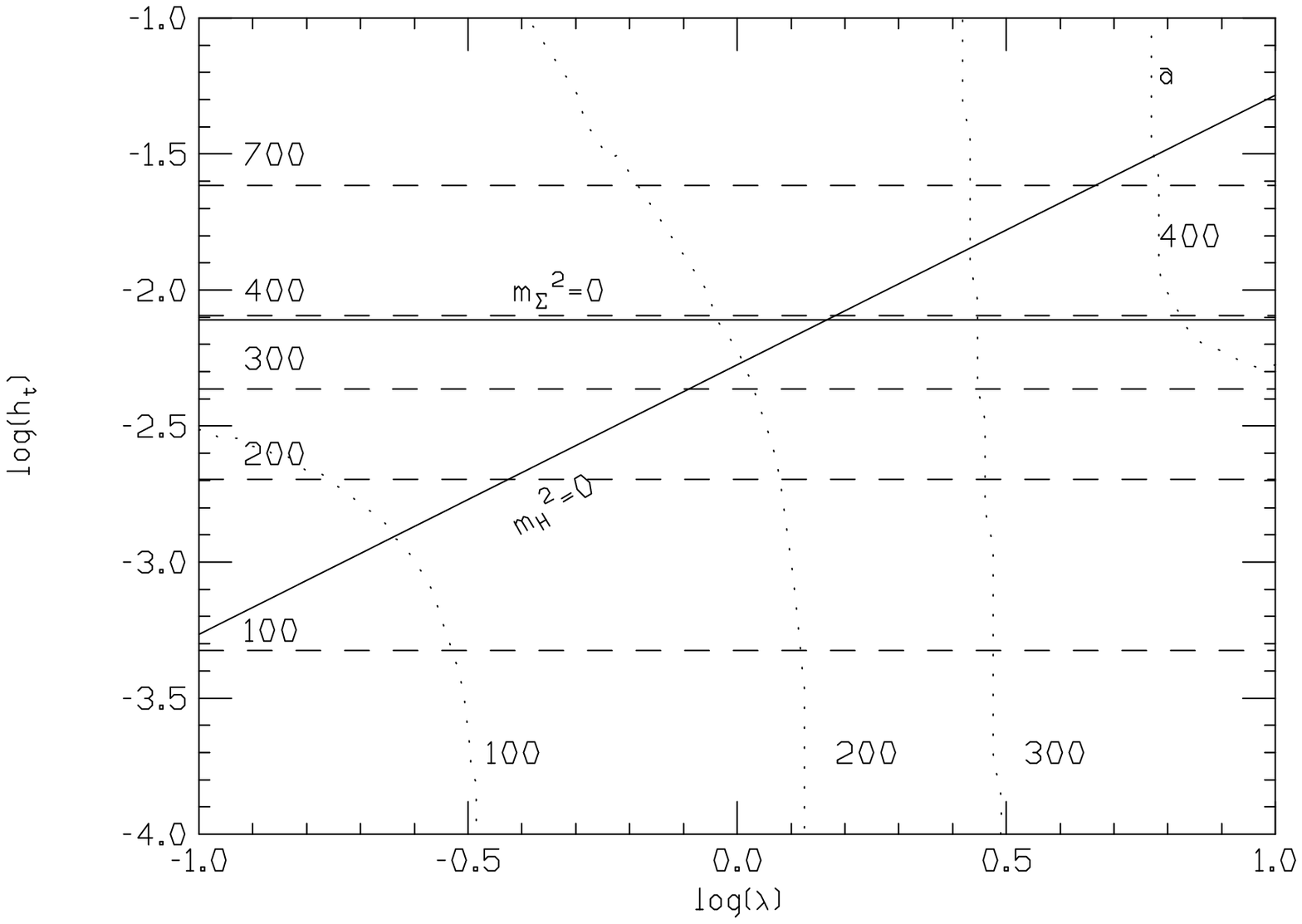}}
\centerline{\epsfysize=2.75 in \epsfbox{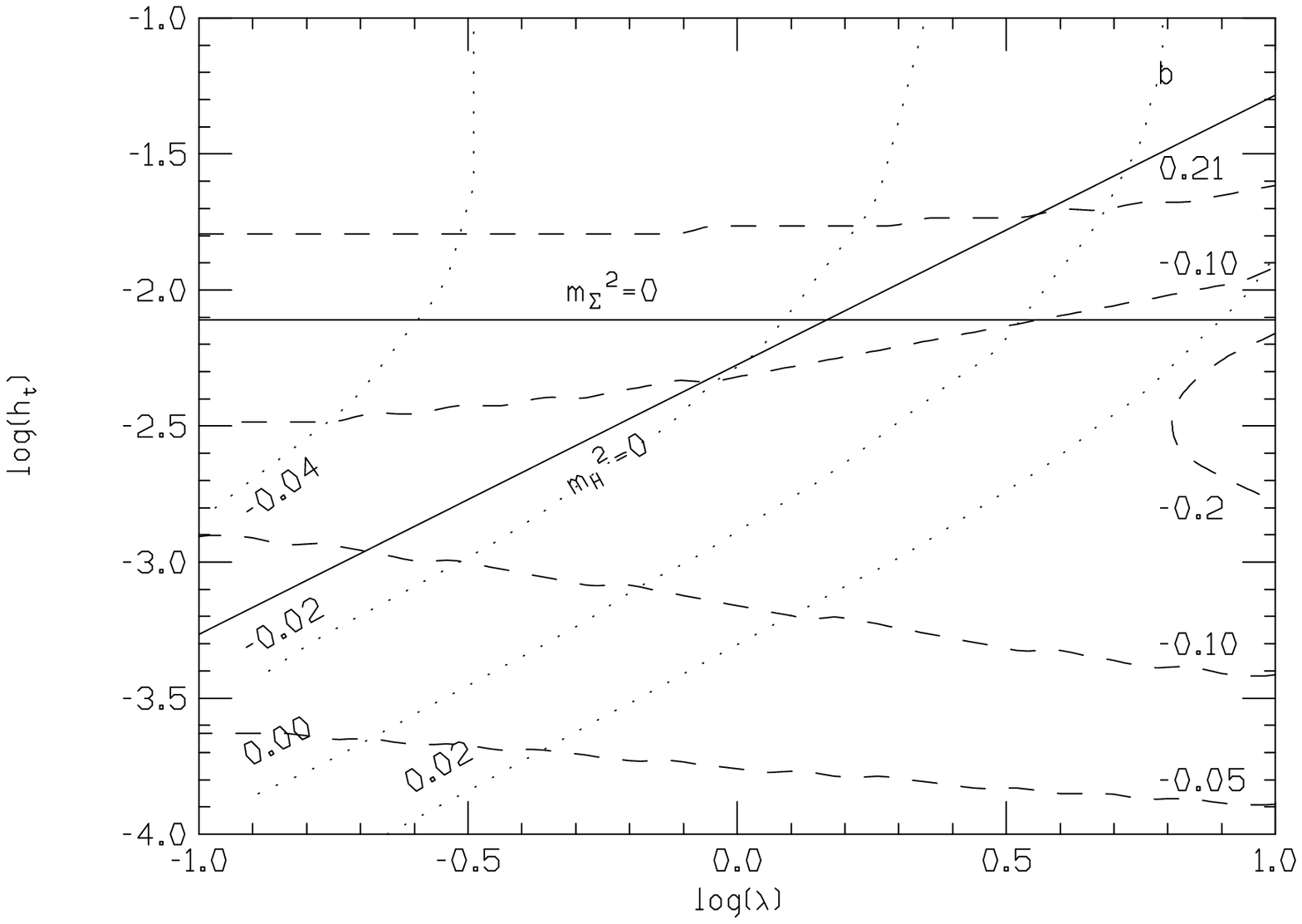}}
\centerline{\epsfysize=2.75 in \epsfbox{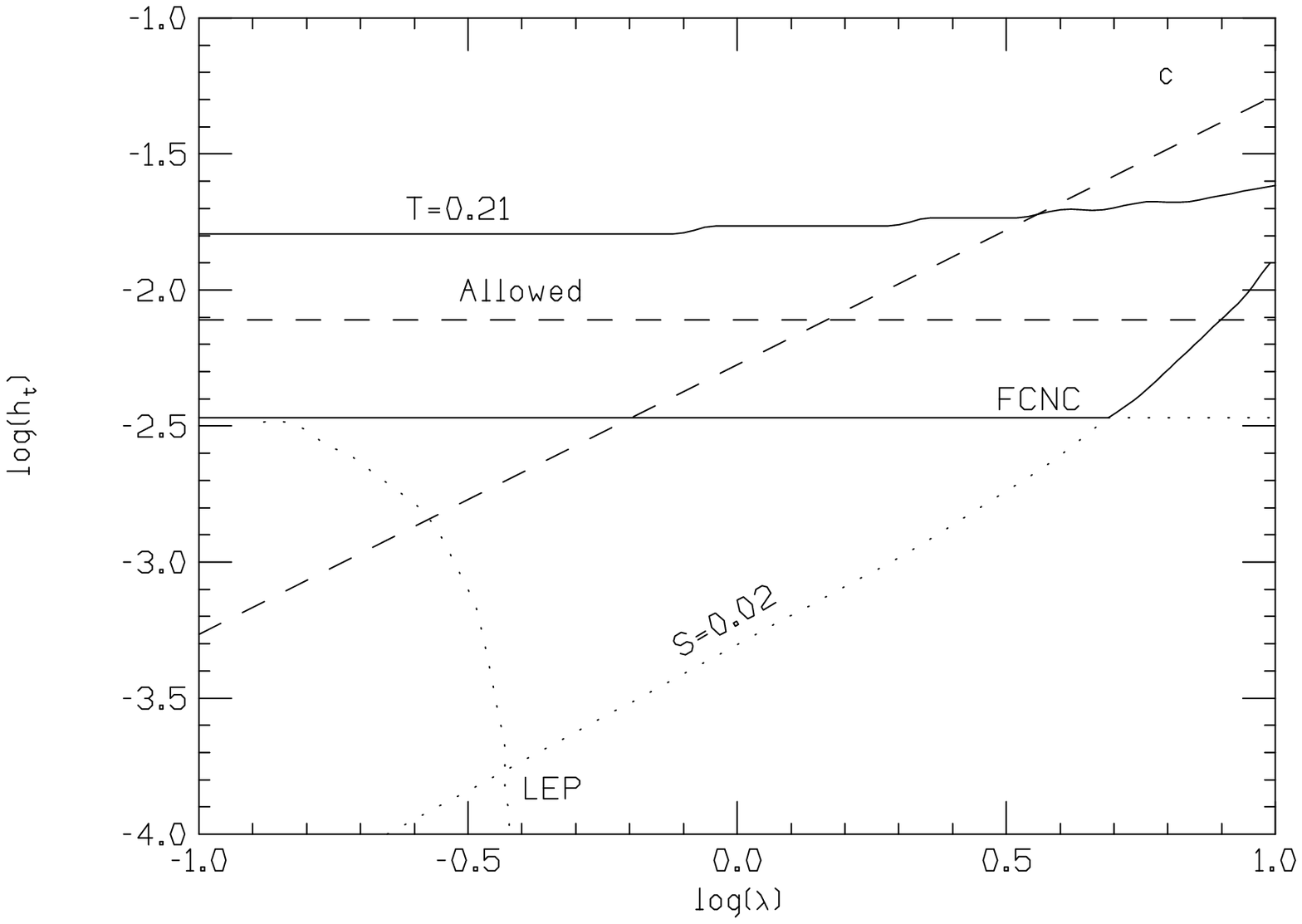}}
 
\caption{Minimal model, $\Lambda=10$~TeV.(a) Neutral (dashed) and
charged (dotted) mass contours, units of GeV, (b) $S$ (dotted) 
and $T$ (dashed) parameter contours, (c) Exclusion regions.}
\label{figure1}
\end{figure}
\begin{figure}  
\centerline{ \epsfysize=2.75 in \epsfbox{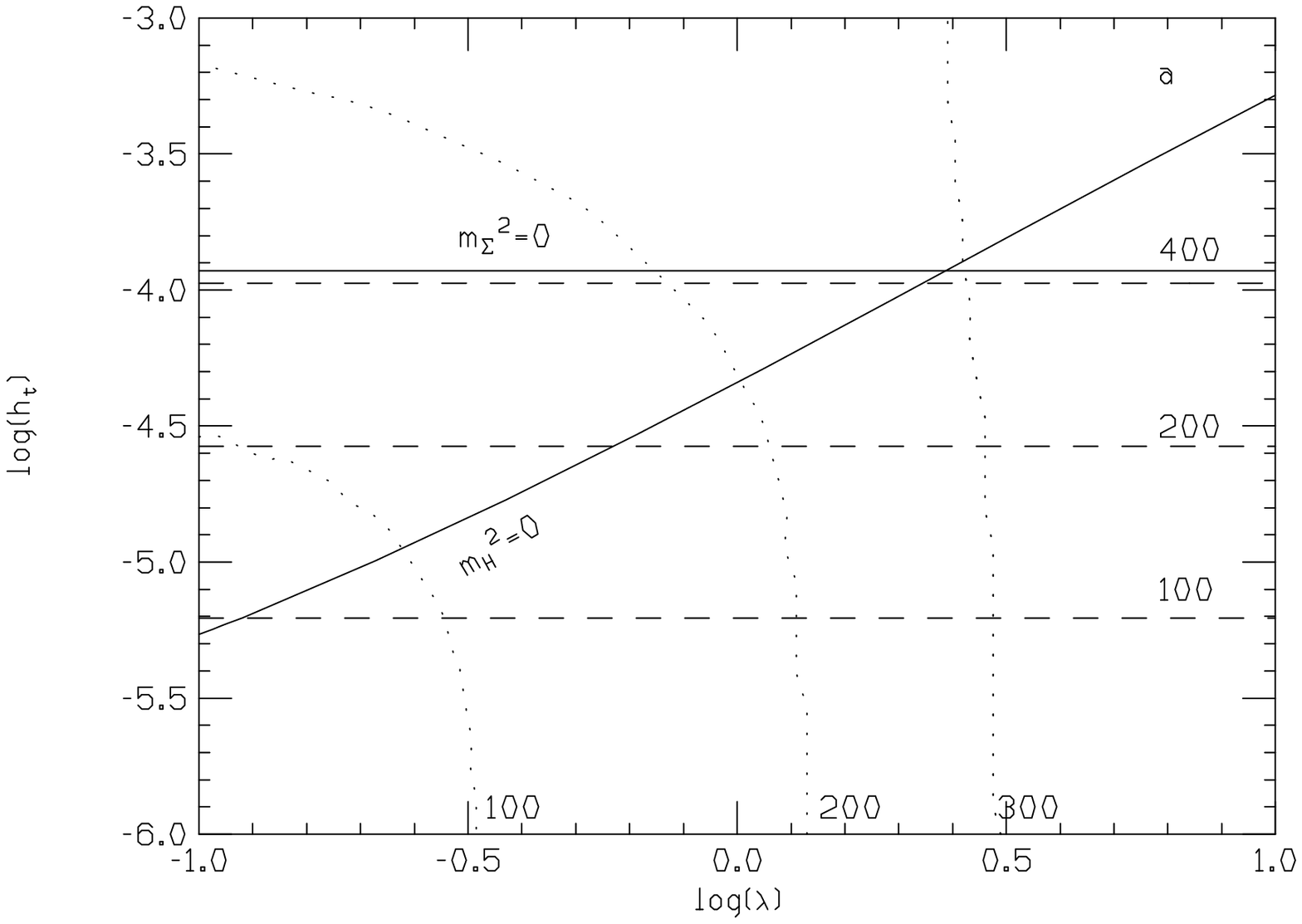}} 
\centerline{ \epsfysize=2.75 in \epsfbox{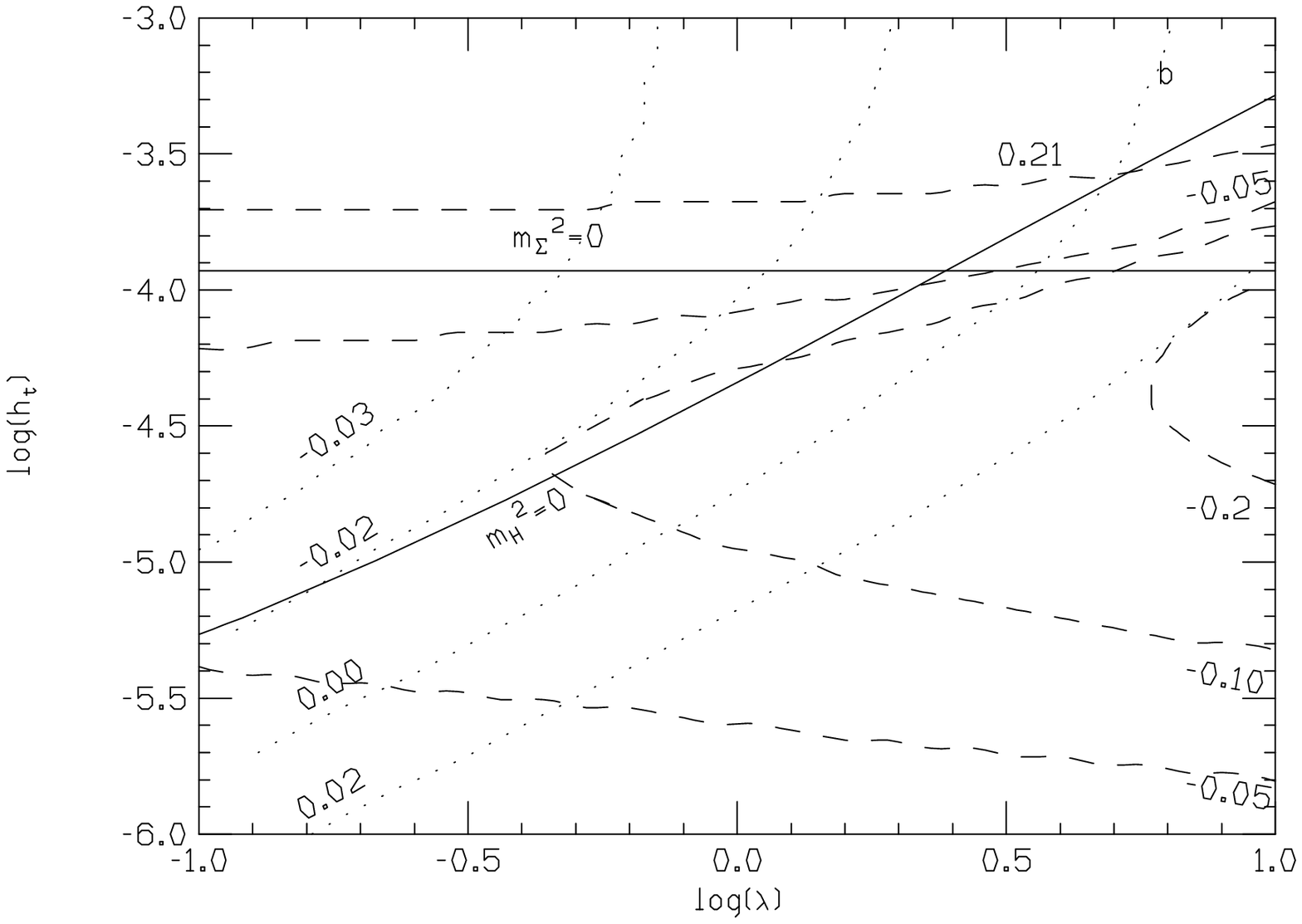}}
\centerline{ \epsfysize=2.75 in \epsfbox{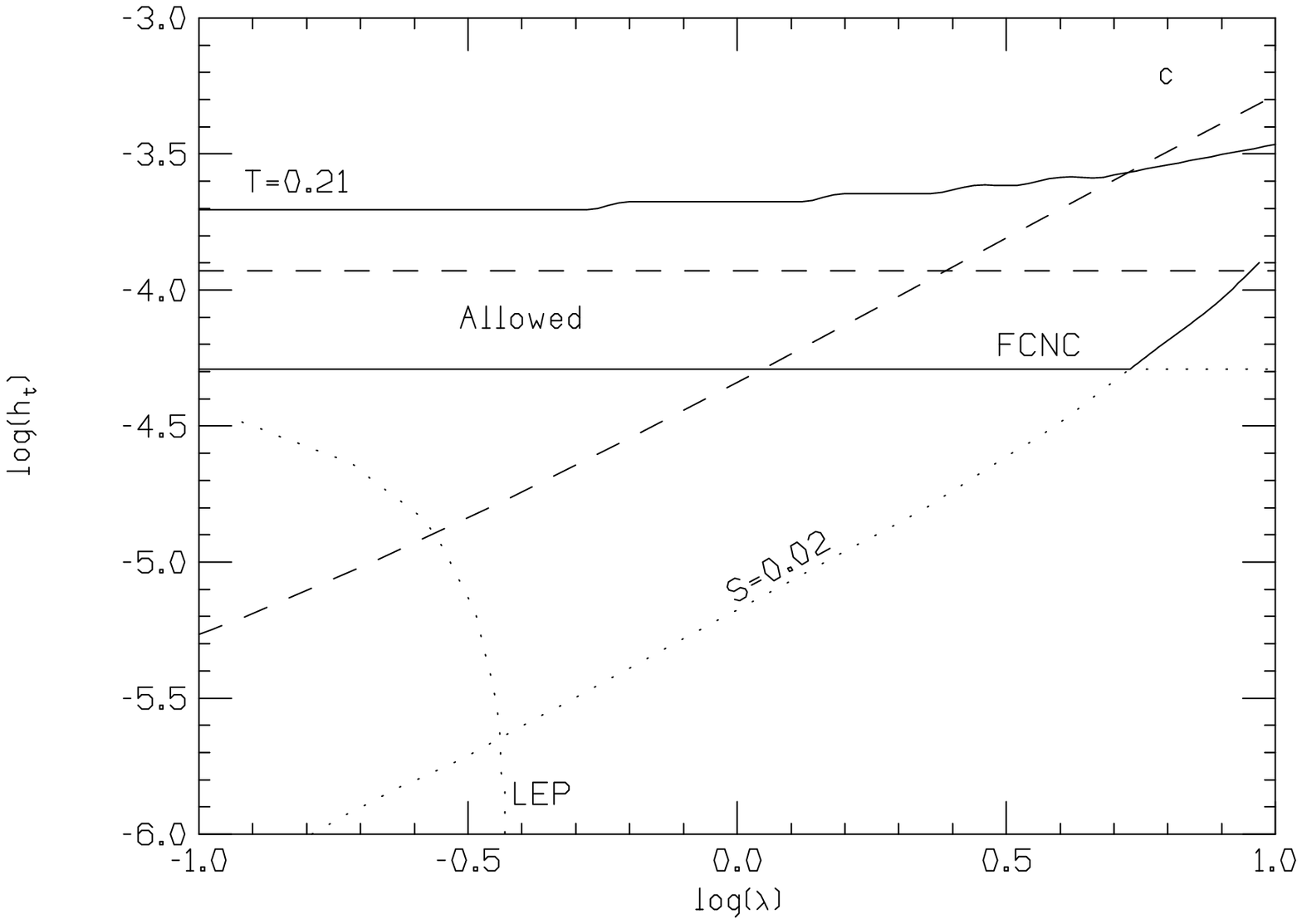}}
 
\caption{Same as Fig.~1, with $\Lambda=100$~TeV.}
\label{figure2}
\end{figure}

\begin{figure}  
\centerline{ \epsfysize=2.75 in \epsfbox{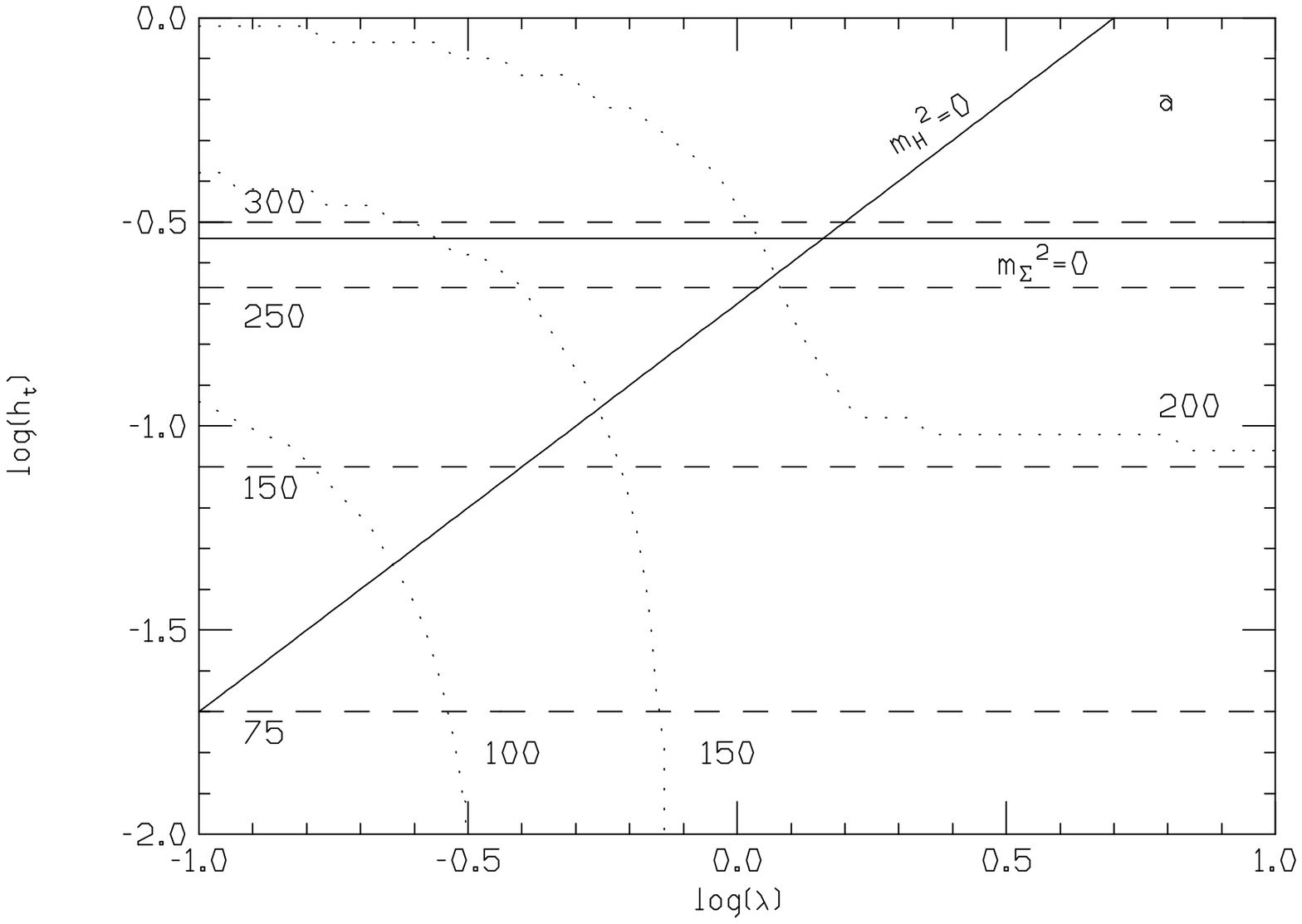}} 
\centerline{ \epsfysize=2.75 in \epsfbox{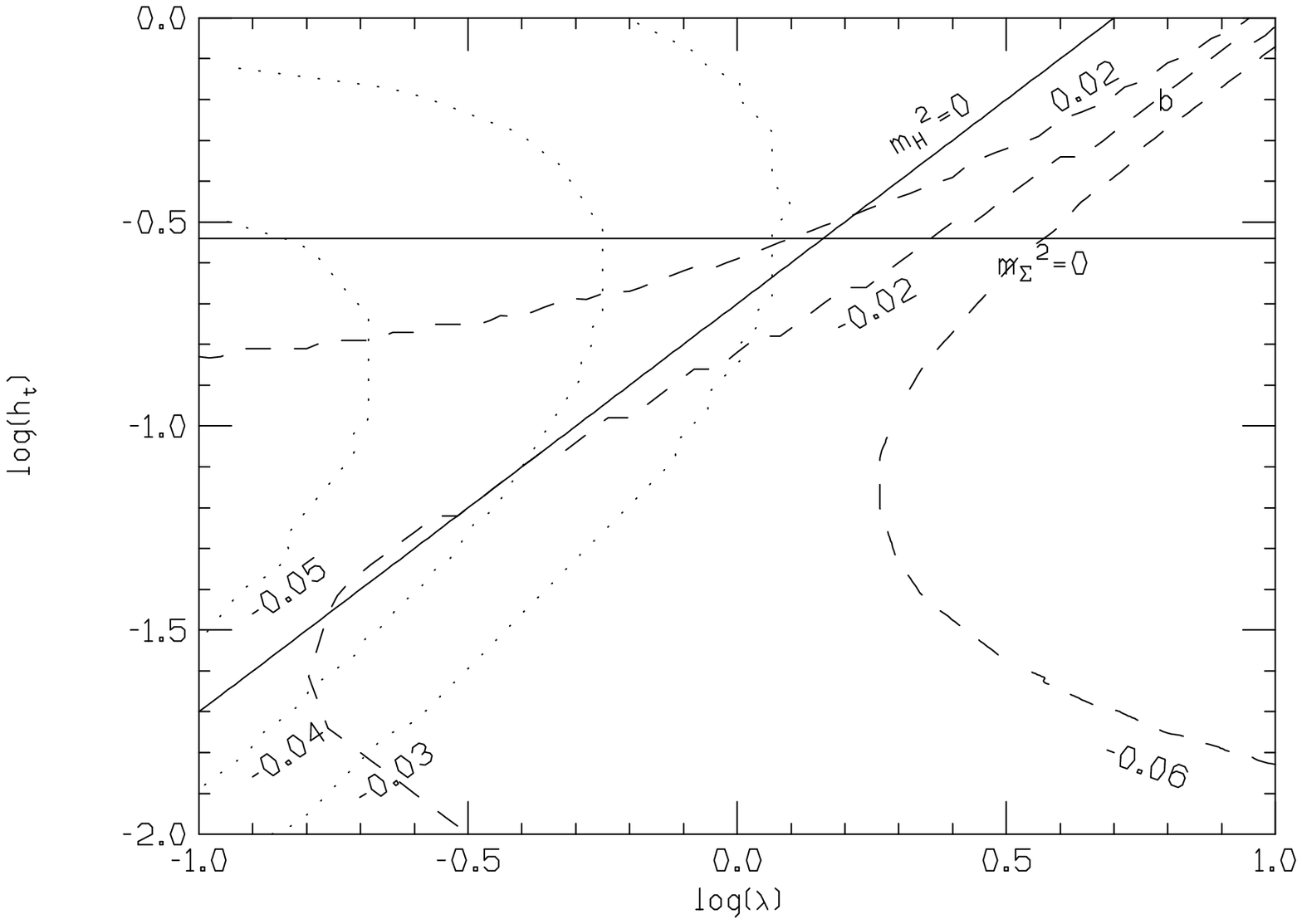}}
\centerline{ \epsfysize=2.75 in \epsfbox{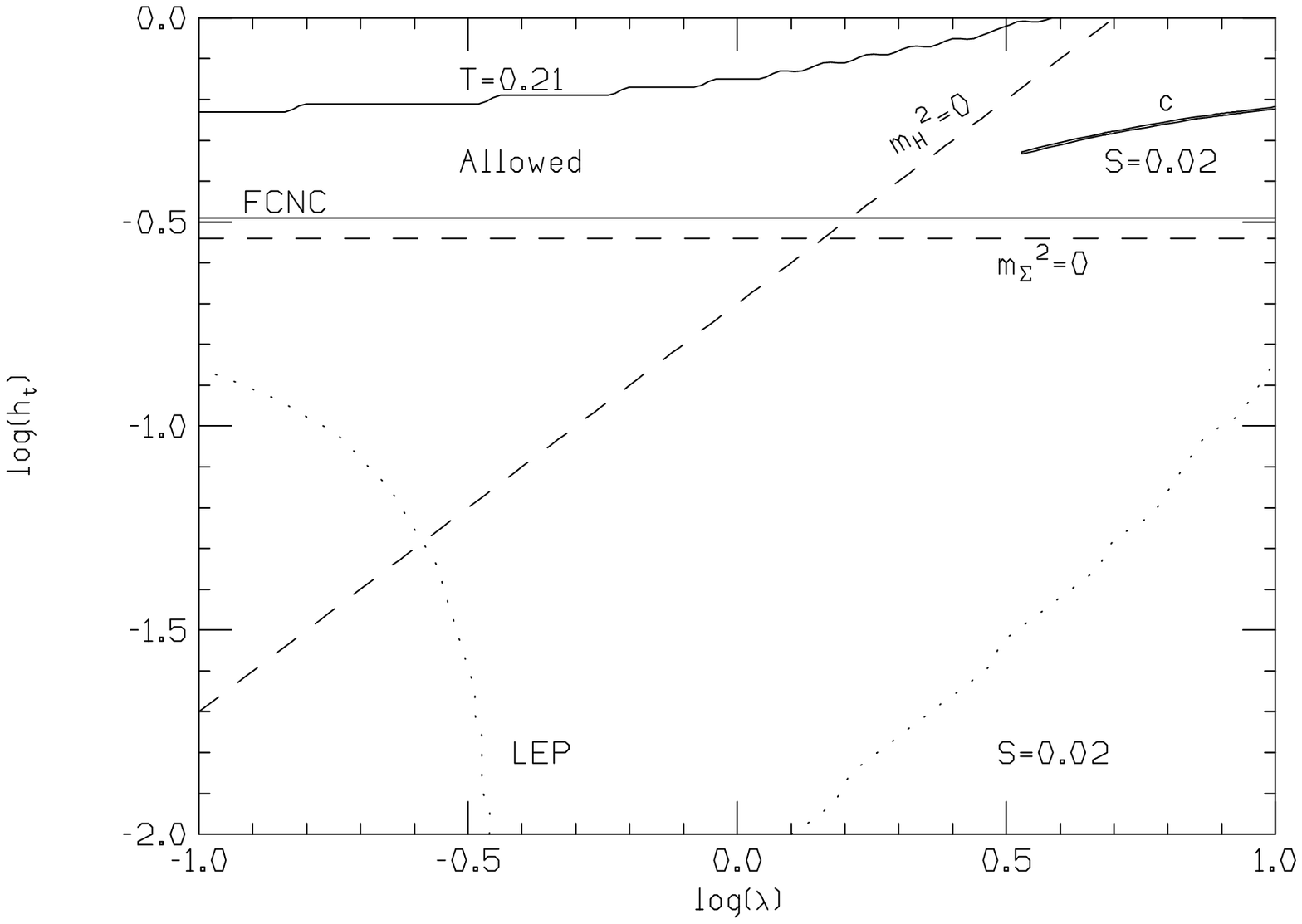}}
 
\caption{General model, $m^2_{H\Sigma}=(400\mbox{ GeV})^2$, $\lambda_0=1$.
Notation is the same as in Figs.~1 and 2.}
\label{figure3}
\end{figure}
%
%%%%%%%%%%%%%%%%%%%%%%%%%%%%%%%%%%%%%%%%%%%%%%%%%%%%%
%  TABLES
%%%%%%%%%%%%%%%%%%%%%%%%%%%%%%%%%%%%%%%%%%%%%%%%%%%%%

\end{document}